\documentclass[aps, pra,twocolumn,showpacs,amsmath,amssymb,superscriptaddress, longbibliography, 10pt]{revtex4-2}

\usepackage{tgtermes}
\usepackage[utf8]{inputenc}
\usepackage{bm}
\usepackage{color}
\usepackage{makeidx}
\usepackage{amsfonts}
\usepackage{amssymb}
\usepackage{amsthm}
\usepackage{graphicx}
\usepackage{epsfig}
\usepackage{subfigure}
\usepackage{amsmath,amssymb}
\usepackage[colorlinks=true,citecolor=blue,urlcolor=blue,linkcolor=blue]{hyperref}
\usepackage{xcolor}
\usepackage{subfigure}

\newcommand{\bea}{\begin{eqnarray}}
\newcommand{\ea}{\end{eqnarray}}
\newcommand{\eea}{\end{eqnarray}}

\newcommand{\sumint}[1]

\begin{document}
	
	%\preprint{APS/123-QED}
\title{Path-dependent correlations in dynamically tuned Ising models and its short-time behavior: application of Magnus expansion}
	
	%\affiliation{%
	%%This line break forced with \textbackslash\textbackslash
	%}%
\author{Xin Wang}
\affiliation{School of Science, Wuhan University of Technology, Wuhan 430070, China}
\author{Bo Yang}
\affiliation{School of Science, Wuhan University of Technology, Wuhan 430070, China}
\affiliation{School of Mathematics and Physics, Hubei Polytechnic University, Huangshi 435003, China}
\author{Bo Zhang}
\email{zhangbo2011@whut.edu.cn}
\affiliation{School of Science, Wuhan University of Technology, Wuhan 430070, China}
\author{Bo Xiong}
\email{boxiong@whut.edu.cn}
\affiliation{School of Science, Wuhan University of Technology, Wuhan 430070, China}

\date{\today}% It is always \today, today,
	%  b%\tableofcontentsut any date may be explicitly specified

\begin{abstract}
We study the buildup of antiferromagnetic (AF) correlation in the dynamically tuned Ising models which are realized by the Rydberg atomic system. In short-time scale, we apply Magnus expansion (ME) to derive the high-order analytic expression of the connected correlation functions and compare it with exactly numerical results for the different lattice geometries, e.g., 1D chain, $2 \times n$ lattice, and $n \times n$ lattice. It is shown that the high-order expansion is required to describe accurately the buildup of AF correlation in the quench dynamics. Moreover, through a 2D square lattice, we find that the magnitude of AF correlation for the same Manhattan distance is proportional to the number of the shortest paths in a sufficiently long time until long and \emph{distinct} paths are involved significantly with the buildup of the correlation. Finally, we propose an applicable experimental setup to realize our findings.	  	
\end{abstract}

%\keywords{}
\maketitle
\section{Introduction}
The study of non-equilibrium dynamics in the strongly correlated system is one of the most challenging areas in quantum physics. If the system is pushed gently away from an equilibrium state, the response of the system can be well understood by the perturbation theory. However, a tough or long-time push can cause some complex effects that are both conceptually and phenomenologically distinct from any behavior presented near equilibrium. These complex phenomena are extremely hard to be captured by both analytic treatment and numerical simulation, eapecially in dimension $d>1$. To develop efficient theory for non-equilibrium dynamics is immediately necessary for us to deeply understand quench dynamics \cite{9,20,10,11}, dynamic phase transition \cite{12}, and resonant excitation dynamics \cite{13}. Moreover, the number of quantum states increases exponentially with atom number and some of these states are highly fragile against environment disturbance, such as quantum and thermal fluctuations. This indicates that to manipulate and measure precisely some quantities associated with the states is extremely difficult, especially in the solid-state system. Thus, it may be applicable to choose other ways, e.g., artificially synthetic quantum system, to explore nonequilibrium dynamics. Recent experimental progress in the Rydberg atomic system provides a versatile quantum simulation platform to study the out-of-equilibrium many-body phenomena owing to its strong and controllable parameters, such as atomic interaction and Rabi frequencies \cite{28,29}. For example, experiments have employed Rydberg atomic systems to investigate many-body scar states \cite{30,18}, phase transition \cite{17,19}, and quench dynamics of AF correlation \cite{3,2,21,22}.

Recently, the realization of the Ising-like model via Rydberg atomic system enables theories and experiments to explore meticulously the buildup of antiferromagnetic correlation. Experiments have shown the Lieb-Robinson bound speed at which the correlation propagate on the range of the interaction is limited \cite{23,24,25,26,27}. Meanwhile the pattern of AF correlation and the effect of some decoherences on such pattern induced by the two-photon excitation scheme and dephasing have been investigated \cite{2}. In the limit of strong dephasing rate where coherence manifested by the off-diagonal elements of the density matrix decays adiabatically, an effective master equation in the diagonal subspace of the density matrix is derived \cite{31} and applied to investigate nonequilibrium dynamics under both the condition of resonant exicitation \cite{32} and the condition of off-resonant excitation \cite{33,34}. For weak dephasing rates, the quench dynamics of spatial correlations in finite 2D Rydberg atomic system are studied by the time-dependent variational principle (TDVP) \cite{35}. Although the effect of the quench speed and various decoherences on the buildup of AF correlation has been discussed intensively, the exact short-time behavior of the nonequilibrium dynamics and the effect of lattice geometries (or naturally different paths) in such system are still not explicit.

In this paper, we employ Magnus method to investigate the short-time behavior of Ising-like model after a linear modulation of the detuning. In a 1D chain, it is shown that the higher-order ME is required to describe accurately the buildup of AF correlation in the quench dynamics. Furthermore, in $2 \times n$ lattice geometry, the connected correlation function is path-dependent and fulfills algebra law; namely the magnitude of the correlation function $C_{ij}$ is multiple times of the number of the shortest paths from $i$ to $j$. This algebra law will fail in a relatively long time when the long and \emph{distinct} paths are involved strongly. Such deduction is confirmed in $n \times n$ lattice system. Finally, we propose an experimental scheme to explore the path-dependent correlation.

In what follows, we introduce the Ising-like Hamiltonian realized by the Rydberg atomic system and the experimental derivation of parameters. Also we show the quench process in our system and the local spin-spin correlation function. In Sec.\,\ref{se3}, we show the analytic results from Magnus method and compare its with numerical results via 1D lattice system. Also we extend our discussion to $2 \times n$ and $n \times n$ lattice geometry and demostrate the effect of lattice geometry on the connected correlation function. Finally, experimental proposal and conclusion are given in Sec.\,\ref{se4}.

\section{model and method}
In this part, we first introduce the Ising-like spin Hamiltonian arisen from Rydberg atoms in an optical lattice \cite{3} or optical tweezers \cite{2}. Then we show the quench protocol of the system parameters and the computation of spatial correlations.
\subsection{Ising-like Rydberg system}
Experimentally, alkali atoms are prepared in the hyperfine ground state $\vert\!\!\downarrow\rangle$ and then coupled coherently to a Rydberg state with a two-photon transition of Rabi frequency $\Omega$ and a detuning $\delta$ [see Fig.\,\ref{Fig1}\,(a)]. Such Rydberg atomic system that we consider here is described by the Hamiltonian
\begin{equation}\label{eq1}
H = \frac{\Omega}{2}\sum_{i}\sigma_{i}^{x}-\delta\sum_{i}n_{i}+\sum_{\langle ij \rangle}U_{ij}n_{i}n_{j},
\end{equation}
where $\sigma_{i}^{x}$ is the transition operator in the $i$-th grid between the ground state $\vert\!\!\downarrow\rangle$ and the Rydberg state $\vert\!\!\uparrow\rangle$, i.e.,\\
$\sigma_{i}^{x}=\vert\!\!\uparrow\rangle\langle\downarrow\!\!\vert_{i}+\vert\!\!\downarrow\rangle\langle\uparrow\!\!\vert_{i}$ and  $n_{i}=\vert\!\!\uparrow\rangle\langle\uparrow\!\!\vert_{i}$ is the projector operator of the Rydberg state for atom $i$. The spin coupling is essentially arisen from the van der Waals interaction between atoms in the Rydberg state, i.e., $U_{ij}=C_{6}/r_{ij}^{6}$, where the distance between $i$-th and $j$-th atoms $r_{ij}=\vert{\rm \textbf r}_{i}-{\rm \textbf r}_{j}\vert$. Conventionally, $U_{ij}$ is extremely large when $r_{ij}$ is less than a blockade radius $R_{b}$ \cite{4,5,6,7,8}, indicating that two atoms within blockade radius can not occupy simultaneously Rydberg states. Experiments have realize the nearest-neighbor distance $r_{i,i+1}\sim R_{b}$, where it is applicable to consider only the nearest-neighbor interaction and for simplification, we set $U_{ij}=U$ for all $\langle ij \rangle$.

In this Ising-like model, one treats approximately $\Omega$ as a transverse field and $\delta$ as a longitudinal field. The asscoiated equilibrium state according to the Ising-like Hamiltonian for $U>0$ displays two phases, paramagnetic (PM) and antiferromagnetic (AF) phases, with a second-order phase transition between them. We prefer to probe the buildup of AF correlation in a quench process, so the initial atoms are prepared in the ground state and the corresponding parameters start from PM regime.

\subsection{Quench process and antiferromagnetic correlation}
\begin{figure}[htpb]
	\centering
	\includegraphics[scale=1.0]{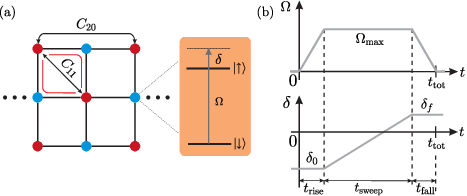}
	\caption{(a) Schematic description of 2D lattice where the Rydberg level system in a single site and the local connected correlation function are displayed. (b) The Rabi frequency $\Omega(t)$ and the detuning $\delta(t)$ are modulated in top and bottom curves, respectively.}
    \label{Fig1}
\end{figure}
To probe the buildup of antiferromagnetic correlation, we employ the quench protocol shown in Fig.\,\ref{Fig1}\,(b) similar to the experiments \cite{2}. Starting with all atoms in the ground state $\vert\!\!\!\downarrow\rangle$, we slowly switch on $\Omega$ with a ramping-up time of $t_{\rm rise}$ until $\Omega=\Omega_{\rm max}$ with a constant detuning of $\delta_{0}$. Afterwards, $\Omega$ is kept at a constant with duration time $t_{\rm sweep}$ and correspondingly $\delta$ linearly increases to $\delta=\delta_{f}$ where the parameters come into the region of the AF phase. Subsequently, $\Omega$ is slowly switched off with a ramping-down time of $t_{\rm fall}$ while $\delta$ is unchanged. To investigate the effect of grid geometry in the AF correlation of the Rydberg atomic system, we measure the connected spin-spin correlation function
\begin{equation}
C_{kl} = \frac{1}{N_{kl}}\sum_{(ij)}\left[\langle n_{i}n_{j}\rangle - \langle n_{i}\rangle \langle n_{j} \rangle\right],
\end{equation} 
where the sum is over pairs $(ij)$ of the same separation ${\rm \textbf r}_{i}-{\rm \textbf r}_{j} = (ka,la)$, and $N_{kl}$ is the number of such atom pairs in the lattice array. Here the grids of the system are distributed uniformly with the separation $a$.

Without loss of generality, we fix the interaction term $U/h=2.7 \rm MHz$ and dynamically tune the detuning $\delta(t)$ and the Rabi frequency $\Omega(t)$ in the experimentally accessible regime \cite{2} for various lattice geometries. The initial state is chosen to be the ground state $\vert\!\!\!\downarrow\rangle$ and the corresponding initial parameters $\delta_{0}/(2\pi) = -6\rm MHz$, $\Omega_{0}/(2\pi) = 0$ for all geometries. Since the parameter regimes of equilibrium AF phase for 1D chain and 2D lattice are different, we set $\Omega_{\rm max}/(2\pi)=0.8\rm MHz$ for 1D chain and $1.8\rm MHz$ for $2 \times n$ and $n \times n$ lattice geometries. To get a starting insight of the development in space and time of the correlations during a ramp, we choose different $\delta_{f}$ under the same $t_{\rm rise}=0.1\mu s$ and $t_{\rm sweep}=0.5\mu s$. After the sweep time, we measure the nearest-neighbor correlation and the next-nearest-neighbor correlation while to explore the relation between $C_{11}$ and $C_{20}$ for 2D lattice, we choose more long sweep time.

\section{result and analysis}\label{se3}
\subsection{Magnus expansion}
At the beginning, we attempt to utilize applicable analytic tools to explore the non-equilibrium dynamics of the Rydberg atomic system. In the short-time scale, the Magnus method (Also Magnus expansion) is a practical way to deal with the non-equilibrium issues. The ME initially is applied to provide an exponential representations of the solution of linear systems of differential equations, especially for varing coefficients. With new techniques involved and original formulation refined, it has been enlarged to explore more complex differential equations, such as, stochastic equations \cite{14}, nonlinear equations \cite{15}, and sturm-Liouville problems \cite{16}.

Our analytic results are based on the Magnus expansion \cite{1} of time propagator. The standard procedue starts to solve the time-dependent Schr\"odinger equation according to Hamiltonian (\ref{eq1}), and then obtains the full many-body propagator $\hat U(T)$ as well as the associated wave function at any time $T$, $\vert \psi(T)\rangle = \hat U(T)\vert\psi(0)\rangle$. The time propagator yields a matrix exponential
\begin{equation}\label{eq3}
\hat U(T) = {\rm exp}\left[-iT\bar H(T)\right],
\end{equation}
where $\bar H(T)$ is a series expansion for the matrix associated to nested commutators of the time-dependent Hamiltonian (\ref{eq1}), i.e., $\bar H(T) = \sum_{k=1}^{\infty}\bar H_{k}(T)$.

Owing to the complexity of many-body Ising model, here we consider the Magnus expansion up to second-order terms and truncate higher-order expansion
\begin{equation}
\bar H_{1} = \frac{1}{T}\int_{0}^{T}H(t_{1})dt_{1},
\end{equation}
\begin{equation}\label{eq5}
\bar H_{2} = \frac{i}{2T}\int_{0}^{T}\left[\int_{0}^{t_{1}}H(t_{2})dt_{2},H(t_{1})\right]dt_{1},
\end{equation}
where the communication of operators fulfills $\left[ \hat A, \hat B\right] = \hat A\hat B - \hat B\hat A$.

In a linear modulation of the detuning as has realized in the experiments \cite{2}, $\delta(t)=\frac{\delta_{f}-\delta_{0}}{T}t+\delta_{0}$, where $\delta_{0}$ is the original
value and $\delta_{f}$ is the final value of $\delta$. One can deduce straightforwardly
\begin{equation}
\bar H_{1} = \frac{\Omega}{2}\sum_{i}\sigma_{i}^{x} - \delta_{\rm avg}\sum_{i}n_{i}+U\sum_{\langle ij\rangle}n_{i}n_{j},
\end{equation}
\\and
\begin{equation}\label{eq7}
\bar H_{2} = \frac{\Omega}{24}(\delta_{\rm f}-\delta_{0})T\sum_{i}\sigma_{i}^{y},
\end{equation}
where $\delta_{\rm avg}=\frac{\delta_{0}+\delta_{f}}{2}$. Note that the final term in Eq.\,(\ref{eq7}) is linearly dependent on $T$, which is originated from the second-order ME. Under the condition that $\int_{0}^{T}\Vert H(t)\Vert_{2}<\pi$, where $\Vert H\Vert_{2}$ is the euclidean norm of $H$ defined as the squared root of the largest eigenvalue of positive semi-definite operator $H^{\dagger}H$, we can expand the matrix exponential (\ref{eq3}) into
\begin{equation}\label{eq8}
\hat U(T) = \sum_{i=1}^{\infty}\frac{(-iT)^{n}}{n!}\bar H^{n}(T),
\end{equation}
where the high-power terms in $T$ appear and will contribute in the following correlation function.

In the next, we express the connected correlation function between the atom in $(0, 0)$ and the one in $(ka, la)$ as
\begin{equation}\label{eq9}
\begin{split}
C_{R}(T) & = \langle \psi(T)\vert n_{(0,0)}n_{(k,l)}\vert \psi(T)\rangle \\
& - \langle \psi(T)\vert n_{(0,0)}\vert \psi(T)\rangle\langle \psi(T)\vert n_{(k,l)}\vert \psi(T)\rangle,
\end{split}
\end{equation}\\
where the Manhattan distance $R = \vert k\vert + \vert l\vert$. From (\ref{eq8}) and (\ref{eq9}), one can see that in the treatment of Magnus approach to the Ising-like Hamiltonian, the buildup of the AF correlation characterised by the connected correlation function requires the expansion of exponential matrix in sufficiently high powers where the interaction term plays an significant role in the dynamic process. Meanwhile, the non-zero anticommutation of the Hamiltonian in (\ref{eq5}) means that it may be necessary to consider high-order ME to obtain more accurate results. Since the buildup of dynamic connected correlation is strongly dependent on the Manhattan distance $R$, that is, for a certain value $C_{R}$ the larger $R$ requires the more time, we consider the high-power terms in $T$ under the second-order ME for the nearest-neighbor correlation $(R = 1)$ and the next-nearest-neighbor correlation $(R = 2)$. While for $R > 2$, we consider only the leading ME.

We now collect symbolic expressions for the power series in $T$ of $C_{R}(T)$, according to $\hat U(T)$ in the Magnus expansion form and the initial state with all sites in the atomic ground state. Moreover, we take into account both a single path on a lattice and the effect of the neighor sites. Insert Eq.\,(\ref{eq8}) into Eq.\,(\ref{eq9}), we obtain the connected correlation function for $R = 1$\\
\begin{minipage}[ht]{0.15\linewidth}
\centering
\includegraphics[scale=1.0]{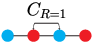}
\end{minipage}
\begin{minipage}[ht]{0.84\linewidth}
\begin{equation}\label{eq10}
C_{R=1}(T) = C_{R=1}^{\rm L,3rd}(T) +C_{R=1}^{\rm S,1st}(T),
\end{equation}
\end{minipage}\\
\\where the first term of the second-order ME
\begin{widetext}
\begin{equation}
C_{R=1}^{\rm S,1st}(T) = -\left[ \frac{1}{20736}+\frac{T^{2}}{5971968}(\delta_{f}-\delta_{0})^{2}\right](\delta_{f}-\delta_{0})^{2}(U-3\delta_{\rm avg})UT^{8}\Omega^{4}.
\end{equation}
\end{widetext}

Correspondingly, the 3rd term of leading-order ME is
\begin{widetext}
\begin{equation}\label{eq12}
\begin{split}
C_{R=1}^{\rm L,3rd}(T) &= C_{R=1}^{\rm L,1st}(T) + \frac{\Omega^{4}T^{8}U}{11520}\left[U^{3}-6U^{2}\delta_{\rm avg}+15U\delta_{\rm avg}^{2}-18\delta_{\rm avg}^{3}+4(U-9\delta_{\rm avg})\Omega^{2}\right]\\
& +\,\frac{\Omega^{4}T^{10}U}{2419200}\left[-3U^{5}+27U^{4}\delta_{\rm avg}-109U^{3}\delta_{\rm avg}^{2}+249U^{2}\delta_{\rm avg}^{3}-337U\delta_{\rm avg}^{4}+255\delta_{\rm avg}^{5}\right.\\
&\left.+\,(275U^{3}-582U^{2}\delta_{\rm avg}-123U\delta_{\rm avg}^{2}+1230\delta_{\rm avg}^{3})\Omega^{2}+(337U+975\delta_{\rm avg})\Omega^{4}\right],
\end{split}
\end{equation}
\end{widetext}
where the lowest power in the leading-order of ME
\begin{equation}\label{eq13}
C_{R=1}^{\rm L,1st}(T) = -\frac{T^{6}}{288}(U^{2}-3U\delta_{\rm avg})\Omega^{4}.
\end{equation}
Note $C_{R=1}^{\rm L,1st}(T)$ is equivalent to the result from a single path for $R=1$ \cite{2}. We stress that since the lowest power in $C_{R=1}^{\rm S,1st}$ is up to $O(T^{10})$, we take the 3rd term of the leading-order ME to match the same order.

In a similar way, we derive the connected correlation function for $R=2$\\
\begin{minipage}[ht]{0.11\linewidth}
\centering
\includegraphics[scale=1.0]{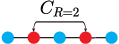}
\end{minipage}
\begin{minipage}[ht]{0.88\linewidth}
\begin{equation}\label{eq14}
C_{R=2}(T) = C_{R=2}^{\rm L,3rd}(T) + C_{R=2}^{\rm S,1st}(T),
\end{equation}
\end{minipage}\\
\\where the second-order contribution
\begin{widetext}
\begin{equation}
C_{R=2}^{\rm S,1st}(T) = \frac{1}{116121600}\left[1+\frac{T^{2}}{144}(\delta_{f}-\delta_{0})^{2}+\frac{T^{4}}{62208}(\delta_{f}-\delta_{0})^{4}\right](\delta_{f}-\delta_{0})^{2}(11U-25\delta_{\rm avg})(7U-15\delta_{\rm avg})U^{2}T^{12}\Omega^{6},
\end{equation}
\end{widetext}
and the high powers in $T$ for the first-order expansion
\begin{widetext}
\begin{equation}\label{eq16}
\begin{split}
C_{R=2}^{\rm L,3rd}(T) & = C_{R=2}^{\rm L,1st}(T) - \frac{\Omega^{6}T^{12}U^{2}}{29030400}\left[52U^{4}-368U^{3}\delta_{\rm avg}+1037U^{2}\delta_{\rm avg}^{2}-1432U\delta_{\rm avg}^{3}+831\delta_{\rm avg}^{4}+4(U^{2}-143U\delta_{\rm avg}\right.\\
&\left.+\,318\delta_{\rm avg}^{2})\Omega^{2}\right]+\frac{\Omega^{6}T^{14}U^{2}}{26824089600}\left[12(121U^{6}-1170U^{5}\delta_{\rm avg}+4952U^{4}\delta_{\rm avg}^{2}-11862U^{3}\delta_{\rm avg}^{3}+17112U^{2}\delta_{\rm avg}^{4}\right.\\
&\left.-\,14322U\delta_{\rm avg}^{5}+5565\delta_{\rm avg}^{6}+(-42405U^{4}+153388U^{3}\delta_{\rm avg}-102973U^{2}\delta_{\rm avg}^{2}-165984U\delta_{avg}^{3}+222390\delta_{\rm avg}^{4})\Omega^{2}\right.\\
&\left.+\,6(-6347U^{2}+7868U\delta_{\rm avg}+25935\delta_{\rm avg}^{2})\Omega^{4}\right],
\end{split}
\end{equation}
\end{widetext}
with
\begin{equation}\label{eq17}
C_{R=2}^{\rm L,1st}(T) = \frac{T^{10}}{2419200}(77U^{4}-340U^{3}\delta_{\rm avg}+375U^{2}\delta_{\rm avg}^{2})\Omega^{6}.
\end{equation}
The next-next-nearest neighbor correlation yields.\\
\begin{minipage}[ht]{0.15\linewidth}
\centering
\includegraphics[scale=1.0]{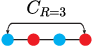}
\end{minipage}
\begin{minipage}[ht]{0.84\linewidth}
\begin{equation}
C_{R=3}(T) = C_{R=3}^{\rm L,1st}(T),
\end{equation}
\end{minipage}\\
\\where
\begin{widetext}
\begin{equation}
C_{R=3}^{\rm L,1st}(T) = -\frac{\Omega^{8}T^{14}}{26824089600}(4279U^{6}-24766U^{5}\delta_{\rm avg}+46725U^{4}\delta_{\rm avg}^{2}-28350U^{3}\delta_{\rm avg}^{3}).
\end{equation}
\end{widetext}

\subsection{1D chain}
\begin{figure}[htpb]
	\centering
	\includegraphics[scale=1.0]{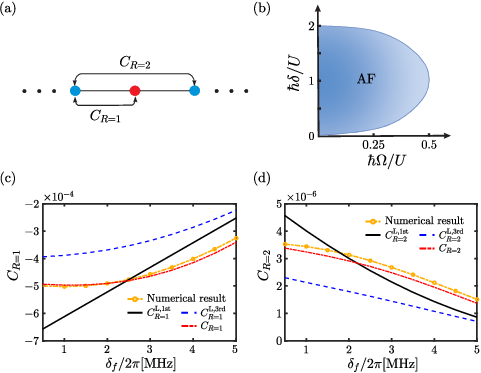}
	\caption{The buildup of antiferromagnetic correlation on 1D system. (a) Schematic description of 1D lattice geometry where we consider a 24-atom chain with periodic boundary conditions. (b) Ground-state phase diagram of the 1D Ising model in Hamiltonian (\ref{eq1}). It displays two phases, one ordered phase, named antiferromagmetic (AF) phase, and paramagnetic (PM) phase, with a second-order phase transition between them. (c) The nearest-neighbor correlation $C_{R=1}$ and (d) the next-nearest-neighbor correlation $C_{R=2}$ as the function of $\delta_{f}$. The yellow dashed line with circle is produced by the results of numerically solving Schr\"odinger equation for Hamiltonian (\ref{eq1}). The black solid line, blue dashed line, and red dotted line show analytic results from the leading ME including the lowest power terms in $T$ and the third lowest power terms, and the second-order ME, respectively.}
    \label{Fig2}
\end{figure}
To compare directly the analytic results from Magnus expansion with exactly numerical solutions, we begin by considering a 1D lattice geometry of 24 Rydberg atoms with periodic boundary conditions and probing the dynamic buildup antiferromagnetic order under various quench processes. The quench protocol is shown in Fig.\,\ref{Fig1}\,(b). For all processes, we choose the identical ramp for $\Omega(t)$, where $\Omega(t)$ increases linearly from $0$ to $\Omega_{\rm max}=2\pi \times 0.8 \rm MHz$ in time interval $t_{\rm rise}=0.1\mu s$, then remains unchanged for $t_{\rm sweep}=0.5\mu s$. Meanwhile we change the different ramping rates for $\delta(t)$ by choosing different final detuning $\delta_{f}$ in a fixed initial detuning $\delta_{0}=-2\pi \times 6 \rm MHz$. Under this modulation, the initial PM state with all atoms occupying $\vert\!\!\downarrow\rangle$ is driven to move toward the Rydberg state $\vert\!\!\uparrow\rangle$ so that the strong interatomic interaction arisen from van der Waals interactions causes the formation of spatial AF structure [see Fig.\,\ref{Fig2}\,(b)]. At $t=t_{\rm rise}+t_{\rm sweep}$, we record the numerical results for $C_{R=1}$ and $C_{R=2}$ and compare its with ME up to different expanding orders. To check the stability of different numerical methods and the effect of finit grid length, we employ both fourth-order Runge Kutta method and time-dependent variational principle (TDVP) method to solve Schr\"odinger equation for Hamiltonian (\ref{eq1}) under different grid lengthes. We find that in the current time scale, the numerical calculation shows similar results of the connected correlation functions for different lattice lengths, e.g., $12$, $16$, $24$.

In Fig.\,\ref{Fig2}\,(c), one can see that the analytic result from Eq.\,(\ref{eq10}) (dotted line) agrees significantly with the exactly numerical solution (dashed line with circles) where with the growth of the final detuning $\delta_{f}$, the nearest-neighbor correlation function $C_{R=1}$ increases approximately in a power-law shape. However, the result from Eq.\,(\ref{eq13}) can not match our numerical result well as $C_{R=1}^{\rm L,1st}$ is completely linear to $\delta_{f}$ (solid line). Although Eq.\,(\ref{eq12}) can describe qualitatively our numerical well (dashed line), there exists a large gap between the analytic result and our simulations, nearly 20\% derivation from the numerical result. Moreover, the similar case appears for the buildup of the AF between the next-nearest-neighbor sites as shown in Fig.\,\ref{Fig2}\,(d): in comparison with Eq.\,(\ref{eq17}) (solid line) and Eq.\,(\ref{eq16}) (dashed line), Eq.\,(\ref{eq14}) (dotted line) provides the results which most closely match the exact numerical results. Thus, we argue that to describe precisely the buildup of AF correlations in the nonequilibrium dynamics with a broad parameter regime, both the third-level powers of Magnus leading expansion and the second-order ME should be taken into account. Also it highlights that the ME in a \emph{single} path can describe well the short-time behavior for the 1D chain.

\subsection{${\textbf2} \times {\rm \textbf n}$ lattice}
\begin{figure}[htpb]
	\centering
	\includegraphics[scale=1.0]{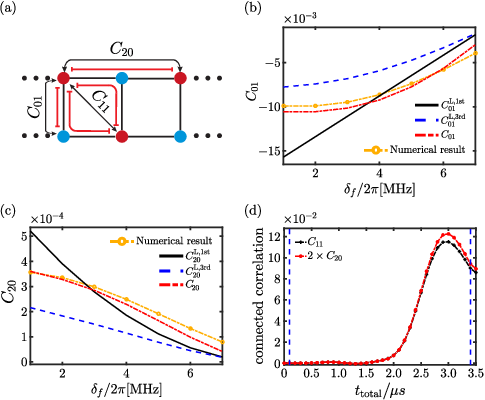}
	\caption{The path-dependent antiferromagnetic correlation on $2 \times n$ lattice. (a) Schematic description of $2 \times n$ lattice system where we consider a $2 \times 12$-atom structure with periodic boundary conditions in row direction. The shortest paths are labeled by the red lines. (b) The nearest-neighbor correlation $C_{01}$ and (c) the next-nearest-neighbor correlation $C_{20}$ as the function of $\delta_{f}$. The production of plots in (b) and (c) are similar to Fig.\,\ref{Fig2}\,(c) and (d). (d) Numerical results of the time evolution of the connected correlation function, $C_{11}$ and $2\times C_{20}$. $\Omega(t)$ and $\delta(t)$ vary under the quench protocol shown in Fig.\,\ref{Fig1}\,(b). The corresponding transition points are labeled by the vertical dashed lines.}
    \label{Fig3}
\end{figure}
To explore the effect of multiple paths on the buildup of antiferromagnetic correlation, we here consider a $2\times n$ lattice geometry with the periodic boundary condition in row direction. The modulation of $\Omega(t)$ and $\delta(t)$ is identical to the one in 1D chain except the maximum value of Rabi frequency $\Omega_{\rm max}=2\pi \times 1.8 \rm MHz$. Also the prepared initial state and the way to record the correlation function are in analogy with 1D case. In contrast to the 1D chain, this lattice geometry has two kinds of the next-nearest-neighbor correlation function, i.e., $C_{11}$ and $C_{20}$ according to the same Manhattan distance $R = 2$ [see Fig.\,\ref{Fig3}\,(a)]. One can easily see that there is only one shortest path for $C_{20}$ while there are two shortest paths for $C_{01}$. To compare the numerical results and the analytic results, we take the single-path correlation $C_{01}$ and $C_{20}$ as an example.

From Fig.\,\ref{Fig3}\,(b) and (c), we obtain similar results as shown in 1D chain: the connected correlation functions including both the third-level powers of Magnus leading expansion and the second-order Magnus expansion, e.g., Eq.\,(\ref{eq10}) and Eq.\,(\ref{eq14}), describes best the numerical results in comparison with the analytic results from the lower order Magnus expansion. Fig.\,\ref{Fig3}\,(d) shows a double relation between two next-nearest-neighbor correlations for a relatively long time, i.e., $C_{11} = 2C_{20}$ for $t < 2.5\mu s$, indicating that the magnitude of the dynamic correlation between spatial two points is proportional to the number of the shortest paths between them. In the many-body language, suppose the dynamic state has the form, $\vert \psi\rangle(t)=\sum_{i}f_{i}(t)\vert i_{A}, i_{p1}, i_{B}, i_{other}\rangle + \sum_{i}f_{i}^{'}(t)\vert i_{A}, i_{p2}, i_{B}, i_{other}\rangle$, where $i_{A}$ and $i_{B}$ are the reference points for measurement. $i_{p1}$ and $i_{p2}$ are the points in the shortest path $1$ and $2$ between $A$ and $B$, respectively, and $i_{other}$ are the other points except the reference points and the points in the shortest path. Suppose all shortest paths are equivalent or $f_{i}(t) = f_{i}^{'}(t)$ for similar basis states. As a result, the corresponding correlation function $C_{AB}=\langle\psi(t)\vert\hat n_{A} \hat n_{B}\vert\psi(t)\rangle-\langle\psi(t)\vert\hat n_{A}\vert\psi(t)\rangle\langle\psi(t)\vert\hat n_{B}\vert\psi(t)\rangle$ becomes path-dependent; namely more paths means larger correlation function. This path-dependent correlation can be broken when long and distinct paths are involved significantly. But in our system, the double relation between $C_{11}$ and $C_{20}$ remains well for a relatively long time. It highlights that although many long and \emph{distinct} paths may affect the double relation, the contribution of the shortest paths still dominates the buildup of AF in the Ising-like atomic system.

\subsection{${\rm \textbf n}\times {\rm \textbf n}$ lattice}
\begin{figure}[htpb]
	\centering
	\includegraphics[scale=1.0]{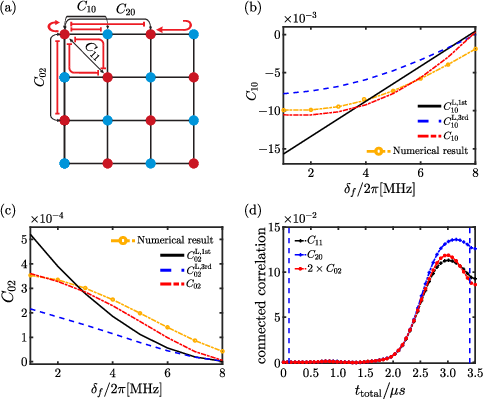}
	\caption{The path-dependent antiferromagnetic correlation on $n \times n$ lattice. (a) Schematic description of $n \times n$ lattice system where we consider a $4\times4$-atom structure with periodic boundary conditions in row direction. (b) The nearest-neighbor correlation $C_{10}$ and (c) the next-nearest-neighbor correlation $C_{02}$ as the function of $\delta_{f}$. (d) Numerical results of the time evolution of the connected correlation function, $C_{11}$, $C_{20}$ and $2 \times C_{02}$.}
    \label{Fig4}
\end{figure}
To verify our theoretical description about the effect of paths on the buildup of correlation, we discuss a $n \times n$ lattice geometry that for two-point correlation generally contains paths more than $2 \times n$ lattice. As a straightforward comparison, we show $4 \times 4$ lattice with the periodic boundary condition in the row direction and modulate the parameters in the same way to the $2 \times n$ lattice. From Fig.\,\ref{Fig4}\,(b) and Fig.\,\ref{Fig4}\,(c), one can see that although more paths are involved with this lattice system, the analytic expression including both the third-level powers of Magnus leading expansion and the second-order ME still work well. This indicates that in the nonequilibrium dynamics with time up to, $T=0.5\mu s$, the single-path ME can evaluate the buildup of AF correlation, even for more complex geometry or essentially more paths involved. Comparing the nearest-neighbor correlation and the next-nearest-neighbor correlation in Fig.\,\ref{Fig2}, Fig.\,\ref{Fig3} and Fig.\,\ref{Fig4}, one can see that the analytic results match best for 1D chain, next for $2 \times n$ lattice, and finally for $n \times n$ lattice, especially for a relatively large $\delta_{f}$. It indicates that with the growth of grid number and dimension, more paths between the reference grids $A$ and $B$ join the buildup of the spin correlation and the single-path ME can not match completely the process despite the single-path contribution still dominates the buildup.

Due to the periodic boundary in row direction, $C_{20}$ has two shortest paths, identical to $C_{11}$, while there exists only one shortest path for $C_{02}$ [see Fig.\,\ref{Fig4}\,(a)]. On basis of the above assumption, for the same Manhattan distance, the magnitude of the connected correlation function is proportional to the number of the shortest paths; namely for $R = 2$ in our system, $C_{20} = C_{11} = 2C_{02}$. Fig.\,\ref{Fig4}\,(d) shows that our numerical results confirm this deduction for $t<2.4\mu s$ and for longer time, the relation is broken, especially between $C_{20}$ and $C_{11}$. It is arisen from the fact that in a long time, only paths in the row contribute most to $C_{20}$ while paths in the row and column contribute nearly equally to $C_{11}$. The similar reason can also answer why $C_{11} = 2C_{02}$ is broken slightly for such a long time.

\section{Discussion and conclusion}\label{se4}
The critical part for experimental realization of path-dependent correlations is to choose an optimal lattice array and reduce considerably the effect of some undesired factors, such as various physical limitations and technical imperfections \cite{36}. We suggest experiments to utilize $2 \times n$ lattice geometry to realize the path-dependent AF correlations because there are less complex paths involved with the buildup of the correlation and the double relation between $C_{11}$ and $C_{20}$ can remain in a longer time in comparison with other lattice geomerties. Experiments have produced various lattice arrays, such as square lattice, triangular lattice \cite{3,2}, Kagome lattice \cite{37,38}, and honeycomb lattice \cite{18}, so it should be applicable to generate $2 \times n$ arrays of optical tweezers. Since our numerical results for $2 \times 6$, $2 \times 8$, $2 \times 12$ are extremely similar, the finite-size effect of the system will be trivial when the size is sufficiently large. Furthermore, to verify the validity of ME for the nearest-neighbor correlation function and the next-nearest-neighbor correlation function, experiments can measure directly $C_{01}$ and $C_{20}$ in this lattice array after a finite sweeping time and compare directly its with analytic expression.

In summary, we investigate the buildup of antiferromagnetic correlations under the quench dynamics of Rydberg atomic system. In a short-time scale, the ME gives a qualitative description for the connected correlation function, which is verified by our exactly numerical solution. Moreover, we find that besides the dependence of the Manhattan distance, the spin correlation function is path-dependent; the magnitude of AF correlation for the same Manhattan distance is proportional to the number of the shortest paths in a long time until long and \emph{distinct} paths take into effect. We argue that the short-time behavior of the AF correlation in the Ising-like system is dominated by the short paths between two reference points and the ME can be extended to more complex lattice geometries (naturally more complex paths). Finally, we propose an optimal lattice array to realize our findings.

\bibliography{Ising}% Produces the bibliography via BibTeX.

\begin{thebibliography}{38}%
\makeatletter
\providecommand \@ifxundefined [1]{%
 \@ifx{#1\undefined}
}%
\providecommand \@ifnum [1]{%
 \ifnum #1\expandafter \@firstoftwo
 \else \expandafter \@secondoftwo
 \fi
}%
\providecommand \@ifx [1]{%
 \ifx #1\expandafter \@firstoftwo
 \else \expandafter \@secondoftwo
 \fi
}%
\providecommand \natexlab [1]{#1}%
\providecommand \enquote  [1]{``#1''}%
\providecommand \bibnamefont  [1]{#1}%
\providecommand \bibfnamefont [1]{#1}%
\providecommand \citenamefont [1]{#1}%
\providecommand \href@noop [0]{\@secondoftwo}%
\providecommand \href [0]{\begingroup \@sanitize@url \@href}%
\providecommand \@href[1]{\@@startlink{#1}\@@href}%
\providecommand \@@href[1]{\endgroup#1\@@endlink}%
\providecommand \@sanitize@url [0]{\catcode `\\12\catcode `\$12\catcode
  `\&12\catcode `\#12\catcode `\^12\catcode `\_12\catcode `\%12\relax}%
\providecommand \@@startlink[1]{}%
\providecommand \@@endlink[0]{}%
\providecommand \url  [0]{\begingroup\@sanitize@url \@url }%
\providecommand \@url [1]{\endgroup\@href {#1}{\urlprefix }}%
\providecommand \urlprefix  [0]{URL }%
\providecommand \Eprint [0]{\href }%
\providecommand \doibase [0]{https://doi.org/}%
\providecommand \selectlanguage [0]{\@gobble}%
\providecommand \bibinfo  [0]{\@secondoftwo}%
\providecommand \bibfield  [0]{\@secondoftwo}%
\providecommand \translation [1]{[#1]}%
\providecommand \BibitemOpen [0]{}%
\providecommand \bibitemStop [0]{}%
\providecommand \bibitemNoStop [0]{.\EOS\space}%
\providecommand \EOS [0]{\spacefactor3000\relax}%
\providecommand \BibitemShut  [1]{\csname bibitem#1\endcsname}%
\let\auto@bib@innerbib\@empty
%</preamble>
\bibitem [{\citenamefont {Sengupta}\ \emph {et~al.}(2004)\citenamefont
  {Sengupta}, \citenamefont {Powell},\ and\ \citenamefont {Sachdev}}]{9}%
  \BibitemOpen
  \bibfield  {author} {\bibinfo {author} {\bibfnamefont {K.}~\bibnamefont
  {Sengupta}}, \bibinfo {author} {\bibfnamefont {S.}~\bibnamefont {Powell}},\
  and\ \bibinfo {author} {\bibfnamefont {S.}~\bibnamefont {Sachdev}},\
  }\bibfield  {title} {\bibinfo {title} {Quench dynamics across quantum
  critical points},\ }\href {https://doi.org/10.1103/PhysRevA.69.053616}
  {\bibfield  {journal} {\bibinfo  {journal} {Phys. Rev. A}\ }\textbf {\bibinfo
  {volume} {69}},\ \bibinfo {pages} {053616} (\bibinfo {year}
  {2004})}\BibitemShut {NoStop}%
\bibitem [{\citenamefont {Calabrese}\ and\ \citenamefont {Cardy}(2006)}]{20}%
  \BibitemOpen
  \bibfield  {author} {\bibinfo {author} {\bibfnamefont {P.}~\bibnamefont
  {Calabrese}}\ and\ \bibinfo {author} {\bibfnamefont {J.}~\bibnamefont
  {Cardy}},\ }\bibfield  {title} {\bibinfo {title} {Time dependence of
  correlation functions following a quantum quench},\ }\href
  {https://doi.org/10.1103/PhysRevLett.96.136801} {\bibfield  {journal}
  {\bibinfo  {journal} {Phys. Rev. Lett.}\ }\textbf {\bibinfo {volume} {96}},\
  \bibinfo {pages} {136801} (\bibinfo {year} {2006})}\BibitemShut {NoStop}%
\bibitem [{\citenamefont {Calabrese}\ \emph {et~al.}(2011)\citenamefont
  {Calabrese}, \citenamefont {Essler},\ and\ \citenamefont {Fagotti}}]{10}%
  \BibitemOpen
  \bibfield  {author} {\bibinfo {author} {\bibfnamefont {P.}~\bibnamefont
  {Calabrese}}, \bibinfo {author} {\bibfnamefont {F.~H.~L.}\ \bibnamefont
  {Essler}},\ and\ \bibinfo {author} {\bibfnamefont {M.}~\bibnamefont
  {Fagotti}},\ }\bibfield  {title} {\bibinfo {title} {Quantum quench in the
  transverse-field ising chain},\ }\href
  {https://doi.org/10.1103/PhysRevLett.106.227203} {\bibfield  {journal}
  {\bibinfo  {journal} {Phys. Rev. Lett.}\ }\textbf {\bibinfo {volume} {106}},\
  \bibinfo {pages} {227203} (\bibinfo {year} {2011})}\BibitemShut {NoStop}%
\bibitem [{\citenamefont {Abeling}\ and\ \citenamefont {Kehrein}(2016)}]{11}%
  \BibitemOpen
  \bibfield  {author} {\bibinfo {author} {\bibfnamefont {N.~O.}\ \bibnamefont
  {Abeling}}\ and\ \bibinfo {author} {\bibfnamefont {S.}~\bibnamefont
  {Kehrein}},\ }\bibfield  {title} {\bibinfo {title} {Quantum quench dynamics
  in the transverse field ising model at nonzero temperatures},\ }\href
  {https://doi.org/10.1103/PhysRevB.93.104302} {\bibfield  {journal} {\bibinfo
  {journal} {Phys. Rev. B}\ }\textbf {\bibinfo {volume} {93}},\ \bibinfo
  {pages} {104302} (\bibinfo {year} {2016})}\BibitemShut {NoStop}%
\bibitem [{\citenamefont {Zhang}\ \emph {et~al.}(2017)\citenamefont {Zhang},
  \citenamefont {Pagano}, \citenamefont {Hess}, \citenamefont {Kyprianidis},
  \citenamefont {Ecker}, \citenamefont {Kaplan}, \citenamefont {Gorshkov},
  \citenamefont {Gong},\ and\ \citenamefont {Monroe}}]{12}%
  \BibitemOpen
  \bibfield  {author} {\bibinfo {author} {\bibfnamefont {J.}~\bibnamefont
  {Zhang}}, \bibinfo {author} {\bibfnamefont {G.}~\bibnamefont {Pagano}},
  \bibinfo {author} {\bibfnamefont {P.~W.}\ \bibnamefont {Hess}}, \bibinfo
  {author} {\bibfnamefont {A.}~\bibnamefont {Kyprianidis}}, \bibinfo {author}
  {\bibfnamefont {P.~B.}\ \bibnamefont {Ecker}}, \bibinfo {author}
  {\bibfnamefont {H.}~\bibnamefont {Kaplan}}, \bibinfo {author} {\bibfnamefont
  {A.~V.}\ \bibnamefont {Gorshkov}}, \bibinfo {author} {\bibfnamefont {Z.~X.}\
  \bibnamefont {Gong}},\ and\ \bibinfo {author} {\bibfnamefont
  {C.}~\bibnamefont {Monroe}},\ }\bibfield  {title} {\bibinfo {title}
  {Observation of a many-body dynamical phase transition with a 53-qubit
  quantum simulator},\ }\href {https://doi.org/10.1038/nature24654} {\bibfield
  {journal} {\bibinfo  {journal} {Nature (London)}\ }\textbf {\bibinfo {volume}
  {551}},\ \bibinfo {pages} {601} (\bibinfo {year} {2017})}\BibitemShut
  {NoStop}%
\bibitem [{\citenamefont {Barredo}\ \emph {et~al.}(2015)\citenamefont
  {Barredo}, \citenamefont {Labuhn}, \citenamefont {Ravets}, \citenamefont
  {Lahaye}, \citenamefont {Browaeys},\ and\ \citenamefont {Adams}}]{13}%
  \BibitemOpen
  \bibfield  {author} {\bibinfo {author} {\bibfnamefont {D.}~\bibnamefont
  {Barredo}}, \bibinfo {author} {\bibfnamefont {H.}~\bibnamefont {Labuhn}},
  \bibinfo {author} {\bibfnamefont {S.}~\bibnamefont {Ravets}}, \bibinfo
  {author} {\bibfnamefont {T.}~\bibnamefont {Lahaye}}, \bibinfo {author}
  {\bibfnamefont {A.}~\bibnamefont {Browaeys}},\ and\ \bibinfo {author}
  {\bibfnamefont {C.~S.}\ \bibnamefont {Adams}},\ }\bibfield  {title} {\bibinfo
  {title} {Coherent excitation transfer in a spin chain of three rydberg
  atoms},\ }\href {https://doi.org/10.1103/PhysRevLett.114.113002} {\bibfield
  {journal} {\bibinfo  {journal} {Phys. Rev. Lett.}\ }\textbf {\bibinfo
  {volume} {114}},\ \bibinfo {pages} {113002} (\bibinfo {year}
  {2015})}\BibitemShut {NoStop}%
\bibitem [{\citenamefont {Labuhn}\ \emph {et~al.}(2016)\citenamefont {Labuhn},
  \citenamefont {Barredo}, \citenamefont {Ravets}, \citenamefont
  {de~L\'es\'eleuc}, \citenamefont {Macr\`i}, \citenamefont {Lahaye},\ and\
  \citenamefont {Browaeys}}]{28}%
  \BibitemOpen
  \bibfield  {author} {\bibinfo {author} {\bibfnamefont {H.}~\bibnamefont
  {Labuhn}}, \bibinfo {author} {\bibfnamefont {D.}~\bibnamefont {Barredo}},
  \bibinfo {author} {\bibfnamefont {S.}~\bibnamefont {Ravets}}, \bibinfo
  {author} {\bibfnamefont {S.}~\bibnamefont {de~L\'es\'eleuc}}, \bibinfo
  {author} {\bibfnamefont {T.}~\bibnamefont {Macr\`i}}, \bibinfo {author}
  {\bibfnamefont {T.}~\bibnamefont {Lahaye}},\ and\ \bibinfo {author}
  {\bibfnamefont {A.}~\bibnamefont {Browaeys}},\ }\bibfield  {title} {\bibinfo
  {title} {Tunable two-dimensional arrays of single rydberg atoms for realizing
  quantum ising models},\ }\href {https://doi.org/10.1038/nature18274}
  {\bibfield  {journal} {\bibinfo  {journal} {Nature (London)}\ }\textbf
  {\bibinfo {volume} {534}},\ \bibinfo {pages} {667} (\bibinfo {year}
  {2016})}\BibitemShut {NoStop}%
\bibitem [{\citenamefont {Barredo}\ \emph {et~al.}(2016)\citenamefont
  {Barredo}, \citenamefont {de~L\'es\'eleuc}, \citenamefont {Lienhard},
  \citenamefont {Lahaye},\ and\ \citenamefont {Browaeys}}]{29}%
  \BibitemOpen
  \bibfield  {author} {\bibinfo {author} {\bibfnamefont {D.}~\bibnamefont
  {Barredo}}, \bibinfo {author} {\bibfnamefont {S.}~\bibnamefont
  {de~L\'es\'eleuc}}, \bibinfo {author} {\bibfnamefont {V.}~\bibnamefont
  {Lienhard}}, \bibinfo {author} {\bibfnamefont {T.}~\bibnamefont {Lahaye}},\
  and\ \bibinfo {author} {\bibfnamefont {A.}~\bibnamefont {Browaeys}},\
  }\bibfield  {title} {\bibinfo {title} {An atom-by-atom assembler of
  defect-free arbitrary two-dimensional atomic arrays},\ }\href
  {https://doi.org/10.1126/science.aah3778} {\bibfield  {journal} {\bibinfo
  {journal} {Science}\ }\textbf {\bibinfo {volume} {354}},\ \bibinfo {pages}
  {1021} (\bibinfo {year} {2016})}\BibitemShut {NoStop}%
\bibitem [{\citenamefont {Turner}\ \emph {et~al.}(2018)\citenamefont {Turner},
  \citenamefont {Michailidis}, \citenamefont {Abanin}, \citenamefont {Serbyn},\
  and\ \citenamefont {Papi{\'c}}}]{30}%
  \BibitemOpen
  \bibfield  {author} {\bibinfo {author} {\bibfnamefont {C.~J.}\ \bibnamefont
  {Turner}}, \bibinfo {author} {\bibfnamefont {A.~A.}\ \bibnamefont
  {Michailidis}}, \bibinfo {author} {\bibfnamefont {D.~A.}\ \bibnamefont
  {Abanin}}, \bibinfo {author} {\bibfnamefont {M.}~\bibnamefont {Serbyn}},\
  and\ \bibinfo {author} {\bibfnamefont {Z.}~\bibnamefont {Papi{\'c}}},\
  }\bibfield  {title} {\bibinfo {title} {Weak ergodicity breaking from quantum
  many-body scars},\ }\href {https://doi.org/10.1038/s41567-018-0137-5}
  {\bibfield  {journal} {\bibinfo  {journal} {Nat. Phys.}\ }\textbf {\bibinfo
  {volume} {14}},\ \bibinfo {pages} {745} (\bibinfo {year} {2018})}\BibitemShut
  {NoStop}%
\bibitem [{\citenamefont {Bluvstein}\ \emph {et~al.}(2021)\citenamefont
  {Bluvstein}, \citenamefont {Omran}, \citenamefont {Levine}, \citenamefont
  {Keesling}, \citenamefont {Semeghini}, \citenamefont {Ebadi}, \citenamefont
  {Wang}, \citenamefont {Michailidis}, \citenamefont {Maskara}, \citenamefont
  {Ho}, \citenamefont {Choi}, \citenamefont {Serbyn}, \citenamefont {Greiner},
  \citenamefont {Vuleti\'c},\ and\ \citenamefont {Lukin}}]{18}%
  \BibitemOpen
  \bibfield  {author} {\bibinfo {author} {\bibfnamefont {D.}~\bibnamefont
  {Bluvstein}}, \bibinfo {author} {\bibfnamefont {A.}~\bibnamefont {Omran}},
  \bibinfo {author} {\bibfnamefont {H.}~\bibnamefont {Levine}}, \bibinfo
  {author} {\bibfnamefont {A.}~\bibnamefont {Keesling}}, \bibinfo {author}
  {\bibfnamefont {G.}~\bibnamefont {Semeghini}}, \bibinfo {author}
  {\bibfnamefont {S.}~\bibnamefont {Ebadi}}, \bibinfo {author} {\bibfnamefont
  {T.~T.}\ \bibnamefont {Wang}}, \bibinfo {author} {\bibfnamefont {A.~A.}\
  \bibnamefont {Michailidis}}, \bibinfo {author} {\bibfnamefont
  {N.}~\bibnamefont {Maskara}}, \bibinfo {author} {\bibfnamefont {W.~W.}\
  \bibnamefont {Ho}}, \bibinfo {author} {\bibfnamefont {S.}~\bibnamefont
  {Choi}}, \bibinfo {author} {\bibfnamefont {M.}~\bibnamefont {Serbyn}},
  \bibinfo {author} {\bibfnamefont {M.}~\bibnamefont {Greiner}}, \bibinfo
  {author} {\bibfnamefont {V.}~\bibnamefont {Vuleti\'c}},\ and\ \bibinfo
  {author} {\bibfnamefont {M.~D.}\ \bibnamefont {Lukin}},\ }\bibfield  {title}
  {\bibinfo {title} {Controlling quantum many-body dynamics in driven rydberg
  atom arrays},\ }\href {https://doi.org/10.1126/science.abg2530} {\bibfield
  {journal} {\bibinfo  {journal} {Science}\ }\textbf {\bibinfo {volume}
  {371}},\ \bibinfo {pages} {1355} (\bibinfo {year} {2021})}\BibitemShut
  {NoStop}%
\bibitem [{\citenamefont {Bernien}\ \emph {et~al.}(2017)\citenamefont
  {Bernien}, \citenamefont {Schwartz}, \citenamefont {Keesling}, \citenamefont
  {H.~Levine}, \citenamefont {Pichler}, \citenamefont {Choi}, \citenamefont
  {Zibrov}, \citenamefont {Endres}, \citenamefont {Greiner}, \citenamefont
  {Vuleti\'c},\ and\ \citenamefont {Lukin}}]{17}%
  \BibitemOpen
  \bibfield  {author} {\bibinfo {author} {\bibfnamefont {H.}~\bibnamefont
  {Bernien}}, \bibinfo {author} {\bibfnamefont {S.}~\bibnamefont {Schwartz}},
  \bibinfo {author} {\bibfnamefont {A.}~\bibnamefont {Keesling}}, \bibinfo
  {author} {\bibfnamefont {A.~O.}\ \bibnamefont {H.~Levine}}, \bibinfo {author}
  {\bibfnamefont {H.}~\bibnamefont {Pichler}}, \bibinfo {author} {\bibfnamefont
  {S.}~\bibnamefont {Choi}}, \bibinfo {author} {\bibfnamefont {A.~S.}\
  \bibnamefont {Zibrov}}, \bibinfo {author} {\bibfnamefont {M.}~\bibnamefont
  {Endres}}, \bibinfo {author} {\bibfnamefont {M.}~\bibnamefont {Greiner}},
  \bibinfo {author} {\bibfnamefont {V.}~\bibnamefont {Vuleti\'c}},\ and\
  \bibinfo {author} {\bibfnamefont {M.~D.}\ \bibnamefont {Lukin}},\ }\bibfield
  {title} {\bibinfo {title} {Probing many-body dynamics on a 51-atom quantum
  simulator},\ }\href {https://doi.org/10.1038/nature24622} {\bibfield
  {journal} {\bibinfo  {journal} {Nature (London)}\ }\textbf {\bibinfo {volume}
  {551}},\ \bibinfo {pages} {579} (\bibinfo {year} {2017})}\BibitemShut
  {NoStop}%
\bibitem [{\citenamefont {Ebadi}\ \emph {et~al.}(2021)\citenamefont {Ebadi},
  \citenamefont {Wang}, \citenamefont {Levine}, \citenamefont {Keesling},
  \citenamefont {Semeghini}, \citenamefont {Omran}, \citenamefont {Bluvstein},
  \citenamefont {Samajdar}, \citenamefont {Pichler}, \citenamefont {Ho},
  \citenamefont {Choi}, \citenamefont {Sachdev}, \citenamefont {Greiner},
  \citenamefont {Vuleti\'c},\ and\ \citenamefont {Lukin}}]{19}%
  \BibitemOpen
  \bibfield  {author} {\bibinfo {author} {\bibfnamefont {S.}~\bibnamefont
  {Ebadi}}, \bibinfo {author} {\bibfnamefont {T.~T.}\ \bibnamefont {Wang}},
  \bibinfo {author} {\bibfnamefont {H.}~\bibnamefont {Levine}}, \bibinfo
  {author} {\bibfnamefont {A.}~\bibnamefont {Keesling}}, \bibinfo {author}
  {\bibfnamefont {G.}~\bibnamefont {Semeghini}}, \bibinfo {author}
  {\bibfnamefont {A.}~\bibnamefont {Omran}}, \bibinfo {author} {\bibfnamefont
  {D.}~\bibnamefont {Bluvstein}}, \bibinfo {author} {\bibfnamefont
  {R.}~\bibnamefont {Samajdar}}, \bibinfo {author} {\bibfnamefont
  {H.}~\bibnamefont {Pichler}}, \bibinfo {author} {\bibfnamefont {W.~W.}\
  \bibnamefont {Ho}}, \bibinfo {author} {\bibfnamefont {S.}~\bibnamefont
  {Choi}}, \bibinfo {author} {\bibfnamefont {S.}~\bibnamefont {Sachdev}},
  \bibinfo {author} {\bibfnamefont {M.}~\bibnamefont {Greiner}}, \bibinfo
  {author} {\bibfnamefont {V.}~\bibnamefont {Vuleti\'c}},\ and\ \bibinfo
  {author} {\bibfnamefont {M.~D.}\ \bibnamefont {Lukin}},\ }\bibfield  {title}
  {\bibinfo {title} {Quantum phases of matter on a 256-atom programmable
  quantum simulator},\ }\href {https://doi.org/10.1038/s41586-021-03582-4}
  {\bibfield  {journal} {\bibinfo  {journal} {Nature (London)}\ }\textbf
  {\bibinfo {volume} {595}},\ \bibinfo {pages} {227} (\bibinfo {year}
  {2021})}\BibitemShut {NoStop}%
\bibitem [{\citenamefont {Guardado-Sanchez}\ \emph {et~al.}(2018)\citenamefont
  {Guardado-Sanchez}, \citenamefont {Brown}, \citenamefont {Mitra},
  \citenamefont {Devakul}, \citenamefont {Huse}, \citenamefont {Schau\ss{}},\
  and\ \citenamefont {Bakr}}]{3}%
  \BibitemOpen
  \bibfield  {author} {\bibinfo {author} {\bibfnamefont {E.}~\bibnamefont
  {Guardado-Sanchez}}, \bibinfo {author} {\bibfnamefont {P.~T.}\ \bibnamefont
  {Brown}}, \bibinfo {author} {\bibfnamefont {D.}~\bibnamefont {Mitra}},
  \bibinfo {author} {\bibfnamefont {T.}~\bibnamefont {Devakul}}, \bibinfo
  {author} {\bibfnamefont {D.~A.}\ \bibnamefont {Huse}}, \bibinfo {author}
  {\bibfnamefont {P.}~\bibnamefont {Schau\ss{}}},\ and\ \bibinfo {author}
  {\bibfnamefont {W.~S.}\ \bibnamefont {Bakr}},\ }\bibfield  {title} {\bibinfo
  {title} {Probing the quench dynamics of antiferromagnetic correlations in a
  2d quantum ising spin system},\ }\href
  {https://doi.org/10.1103/PhysRevX.8.021069} {\bibfield  {journal} {\bibinfo
  {journal} {Phys. Rev. X}\ }\textbf {\bibinfo {volume} {8}},\ \bibinfo {pages}
  {021069} (\bibinfo {year} {2018})}\BibitemShut {NoStop}%
\bibitem [{\citenamefont {Lienhard}\ \emph {et~al.}(2018)\citenamefont
  {Lienhard}, \citenamefont {de~L\'es\'eleuc}, \citenamefont {Barredo},
  \citenamefont {Lahaye}, \citenamefont {Browaeys}, \citenamefont {Schuler},
  \citenamefont {Henry},\ and\ \citenamefont {L\"auchli}}]{2}%
  \BibitemOpen
  \bibfield  {author} {\bibinfo {author} {\bibfnamefont {V.}~\bibnamefont
  {Lienhard}}, \bibinfo {author} {\bibfnamefont {S.}~\bibnamefont
  {de~L\'es\'eleuc}}, \bibinfo {author} {\bibfnamefont {D.}~\bibnamefont
  {Barredo}}, \bibinfo {author} {\bibfnamefont {T.}~\bibnamefont {Lahaye}},
  \bibinfo {author} {\bibfnamefont {A.}~\bibnamefont {Browaeys}}, \bibinfo
  {author} {\bibfnamefont {M.}~\bibnamefont {Schuler}}, \bibinfo {author}
  {\bibfnamefont {L.-P.}\ \bibnamefont {Henry}},\ and\ \bibinfo {author}
  {\bibfnamefont {A.~M.}\ \bibnamefont {L\"auchli}},\ }\bibfield  {title}
  {\bibinfo {title} {Observing the space- and time-dependent growth of
  correlations in dynamically tuned synthetic ising models with
  antiferromagnetic interactions},\ }\href
  {https://doi.org/10.1103/PhysRevX.8.021070} {\bibfield  {journal} {\bibinfo
  {journal} {Phys. Rev. X}\ }\textbf {\bibinfo {volume} {8}},\ \bibinfo {pages}
  {021070} (\bibinfo {year} {2018})}\BibitemShut {NoStop}%
\bibitem [{\citenamefont {Keesling}\ \emph {et~al.}(2019)\citenamefont
  {Keesling}, \citenamefont {Omran}, \citenamefont {Levine}, \citenamefont
  {Bernien}, \citenamefont {Pichler}, \citenamefont {Choi}, \citenamefont
  {Samajdar}, \citenamefont {Schwartz}, \citenamefont {Silvi}, \citenamefont
  {Sachdev}, \citenamefont {Zoller}, \citenamefont {Endres}, \citenamefont
  {Greiner}, \citenamefont {Vuleti\'c},\ and\ \citenamefont {Lukin}}]{21}%
  \BibitemOpen
  \bibfield  {author} {\bibinfo {author} {\bibfnamefont {A.}~\bibnamefont
  {Keesling}}, \bibinfo {author} {\bibfnamefont {A.}~\bibnamefont {Omran}},
  \bibinfo {author} {\bibfnamefont {H.}~\bibnamefont {Levine}}, \bibinfo
  {author} {\bibfnamefont {H.}~\bibnamefont {Bernien}}, \bibinfo {author}
  {\bibfnamefont {H.}~\bibnamefont {Pichler}}, \bibinfo {author} {\bibfnamefont
  {S.}~\bibnamefont {Choi}}, \bibinfo {author} {\bibfnamefont {R.}~\bibnamefont
  {Samajdar}}, \bibinfo {author} {\bibfnamefont {S.}~\bibnamefont {Schwartz}},
  \bibinfo {author} {\bibfnamefont {P.}~\bibnamefont {Silvi}}, \bibinfo
  {author} {\bibfnamefont {S.}~\bibnamefont {Sachdev}}, \bibinfo {author}
  {\bibfnamefont {P.}~\bibnamefont {Zoller}}, \bibinfo {author} {\bibfnamefont
  {M.}~\bibnamefont {Endres}}, \bibinfo {author} {\bibfnamefont
  {M.}~\bibnamefont {Greiner}}, \bibinfo {author} {\bibfnamefont
  {V.}~\bibnamefont {Vuleti\'c}},\ and\ \bibinfo {author} {\bibfnamefont
  {M.~D.}\ \bibnamefont {Lukin}},\ }\bibfield  {title} {\bibinfo {title}
  {Quantum kibble-zurek mechanism and critical dynamics on a programmable
  rydberg simulator},\ }\href {https://doi.org/10.1038/s41586-019-1070-1}
  {\bibfield  {journal} {\bibinfo  {journal} {Nature (London)}\ }\textbf
  {\bibinfo {volume} {568}},\ \bibinfo {pages} {207} (\bibinfo {year}
  {2019})}\BibitemShut {NoStop}%
\bibitem [{\citenamefont {Scholl}\ \emph {et~al.}(2021)\citenamefont {Scholl},
  \citenamefont {Schuler}, \citenamefont {Williams}, \citenamefont
  {Eberharter}, \citenamefont {Barredo}, \citenamefont {Schymik}, \citenamefont
  {Lienhard}, \citenamefont {Henry}, \citenamefont {Lang}, \citenamefont
  {Lahaye}, \citenamefont {L\"auchli},\ and\ \citenamefont {Browaeys}}]{22}%
  \BibitemOpen
  \bibfield  {author} {\bibinfo {author} {\bibfnamefont {P.}~\bibnamefont
  {Scholl}}, \bibinfo {author} {\bibfnamefont {M.}~\bibnamefont {Schuler}},
  \bibinfo {author} {\bibfnamefont {H.~J.}\ \bibnamefont {Williams}}, \bibinfo
  {author} {\bibfnamefont {A.~A.}\ \bibnamefont {Eberharter}}, \bibinfo
  {author} {\bibfnamefont {D.}~\bibnamefont {Barredo}}, \bibinfo {author}
  {\bibfnamefont {K.-N.}\ \bibnamefont {Schymik}}, \bibinfo {author}
  {\bibfnamefont {V.}~\bibnamefont {Lienhard}}, \bibinfo {author}
  {\bibfnamefont {L.-P.}\ \bibnamefont {Henry}}, \bibinfo {author}
  {\bibfnamefont {T.~C.}\ \bibnamefont {Lang}}, \bibinfo {author}
  {\bibfnamefont {T.}~\bibnamefont {Lahaye}}, \bibinfo {author} {\bibfnamefont
  {A.~M.}\ \bibnamefont {L\"auchli}},\ and\ \bibinfo {author} {\bibfnamefont
  {A.}~\bibnamefont {Browaeys}},\ }\bibfield  {title} {\bibinfo {title}
  {Quantum simulation of 2d antiferromagnets with hundreds of rydberg atoms},\
  }\href {https://doi.org/10.1038/s41586-021-03585-1} {\bibfield  {journal}
  {\bibinfo  {journal} {Nature (London)}\ }\textbf {\bibinfo {volume} {595}},\
  \bibinfo {pages} {233} (\bibinfo {year} {2021})}\BibitemShut {NoStop}%
\bibitem [{\citenamefont {Lieb}\ and\ \citenamefont {Robinson}(1972)}]{23}%
  \BibitemOpen
  \bibfield  {author} {\bibinfo {author} {\bibfnamefont {E.~H.}\ \bibnamefont
  {Lieb}}\ and\ \bibinfo {author} {\bibfnamefont {D.~W.}\ \bibnamefont
  {Robinson}},\ }\bibfield  {title} {\bibinfo {title} {The finite group
  velocity of quantum spin systems},\ }\href
  {https://doi.org/10.1007/BF01645779} {\bibfield  {journal} {\bibinfo
  {journal} {Commun. Math. Phys.}\ }\textbf {\bibinfo {volume} {28}},\ \bibinfo
  {pages} {251} (\bibinfo {year} {1972})}\BibitemShut {NoStop}%
\bibitem [{\citenamefont {Bravyi}\ \emph {et~al.}(2006)\citenamefont {Bravyi},
  \citenamefont {Hastings},\ and\ \citenamefont {Verstraete}}]{24}%
  \BibitemOpen
  \bibfield  {author} {\bibinfo {author} {\bibfnamefont {S.}~\bibnamefont
  {Bravyi}}, \bibinfo {author} {\bibfnamefont {M.~B.}\ \bibnamefont
  {Hastings}},\ and\ \bibinfo {author} {\bibfnamefont {F.}~\bibnamefont
  {Verstraete}},\ }\bibfield  {title} {\bibinfo {title} {Lieb-robinson bounds
  and the generation of correlations and topological quantum order},\ }\href
  {https://doi.org/10.1103/PhysRevLett.97.050401} {\bibfield  {journal}
  {\bibinfo  {journal} {Phys. Rev. Lett.}\ }\textbf {\bibinfo {volume} {97}},\
  \bibinfo {pages} {050401} (\bibinfo {year} {2006})}\BibitemShut {NoStop}%
\bibitem [{\citenamefont {Eisert}\ \emph {et~al.}(2013)\citenamefont {Eisert},
  \citenamefont {van~den Worm}, \citenamefont {Manmana},\ and\ \citenamefont
  {Kastner}}]{25}%
  \BibitemOpen
  \bibfield  {author} {\bibinfo {author} {\bibfnamefont {J.}~\bibnamefont
  {Eisert}}, \bibinfo {author} {\bibfnamefont {M.}~\bibnamefont {van~den
  Worm}}, \bibinfo {author} {\bibfnamefont {S.~R.}\ \bibnamefont {Manmana}},\
  and\ \bibinfo {author} {\bibfnamefont {M.}~\bibnamefont {Kastner}},\
  }\bibfield  {title} {\bibinfo {title} {Breakdown of quasilocality in
  long-range quantum lattice models},\ }\href
  {https://doi.org/10.1103/PhysRevLett.111.260401} {\bibfield  {journal}
  {\bibinfo  {journal} {Phys. Rev. Lett.}\ }\textbf {\bibinfo {volume} {111}},\
  \bibinfo {pages} {260401} (\bibinfo {year} {2013})}\BibitemShut {NoStop}%
\bibitem [{\citenamefont {Foss-Feig}\ \emph {et~al.}(2015)\citenamefont
  {Foss-Feig}, \citenamefont {Gong}, \citenamefont {Clark},\ and\ \citenamefont
  {Gorshkov}}]{26}%
  \BibitemOpen
  \bibfield  {author} {\bibinfo {author} {\bibfnamefont {M.}~\bibnamefont
  {Foss-Feig}}, \bibinfo {author} {\bibfnamefont {Z.-X.}\ \bibnamefont {Gong}},
  \bibinfo {author} {\bibfnamefont {C.~W.}\ \bibnamefont {Clark}},\ and\
  \bibinfo {author} {\bibfnamefont {A.~V.}\ \bibnamefont {Gorshkov}},\
  }\bibfield  {title} {\bibinfo {title} {Nearly linear light cones in
  long-range interacting quantum systems},\ }\href
  {https://doi.org/10.1103/PhysRevLett.114.157201} {\bibfield  {journal}
  {\bibinfo  {journal} {Phys. Rev. Lett.}\ }\textbf {\bibinfo {volume} {114}},\
  \bibinfo {pages} {157201} (\bibinfo {year} {2015})}\BibitemShut {NoStop}%
\bibitem [{\citenamefont {Cevolani}\ \emph {et~al.}(2016)\citenamefont
  {Cevolani}, \citenamefont {Carleo},\ and\ \citenamefont
  {Sanchez-Palencia}}]{27}%
  \BibitemOpen
  \bibfield  {author} {\bibinfo {author} {\bibfnamefont {L.}~\bibnamefont
  {Cevolani}}, \bibinfo {author} {\bibfnamefont {G.}~\bibnamefont {Carleo}},\
  and\ \bibinfo {author} {\bibfnamefont {L.}~\bibnamefont {Sanchez-Palencia}},\
  }\bibfield  {title} {\bibinfo {title} {Spreading of correlations in exactly
  solvable quantum models with long-range interactions in arbitrary
  dimensions},\ }\href {https://doi.org/10.1088/1367-2630/18/9/093002}
  {\bibfield  {journal} {\bibinfo  {journal} {New J. Phys.}\ }\textbf {\bibinfo
  {volume} {18}},\ \bibinfo {pages} {093002} (\bibinfo {year}
  {2016})}\BibitemShut {NoStop}%
\bibitem [{\citenamefont {Marcuzzi}\ \emph {et~al.}(2014)\citenamefont
  {Marcuzzi}, \citenamefont {Schick}, \citenamefont {Olmos},\ and\
  \citenamefont {Lesanovsky}}]{31}%
  \BibitemOpen
  \bibfield  {author} {\bibinfo {author} {\bibfnamefont {M.}~\bibnamefont
  {Marcuzzi}}, \bibinfo {author} {\bibfnamefont {J.}~\bibnamefont {Schick}},
  \bibinfo {author} {\bibfnamefont {B.}~\bibnamefont {Olmos}},\ and\ \bibinfo
  {author} {\bibfnamefont {I.}~\bibnamefont {Lesanovsky}},\ }\bibfield  {title}
  {\bibinfo {title} {Effective dynamics of strongly dissipative rydberg
  gases},\ }\href {https://doi.org/10.1088/1751-8113/47/48/482001} {\bibfield
  {journal} {\bibinfo  {journal} {J. Phys. A: Math. Theor.}\ }\textbf {\bibinfo
  {volume} {47}},\ \bibinfo {pages} {482001} (\bibinfo {year}
  {2014})}\BibitemShut {NoStop}%
\bibitem [{\citenamefont {Lesanovsky}\ and\ \citenamefont
  {Garrahan}(2013)}]{32}%
  \BibitemOpen
  \bibfield  {author} {\bibinfo {author} {\bibfnamefont {I.}~\bibnamefont
  {Lesanovsky}}\ and\ \bibinfo {author} {\bibfnamefont {J.~P.}\ \bibnamefont
  {Garrahan}},\ }\bibfield  {title} {\bibinfo {title} {Kinetic constraints,
  hierarchical relaxation, and onset of glassiness in strongly interacting and
  dissipative rydberg gases},\ }\href
  {https://doi.org/10.1103/PhysRevLett.111.215305} {\bibfield  {journal}
  {\bibinfo  {journal} {Phys. Rev. Lett.}\ }\textbf {\bibinfo {volume} {111}},\
  \bibinfo {pages} {215305} (\bibinfo {year} {2013})}\BibitemShut {NoStop}%
\bibitem [{\citenamefont {Lesanovsky}\ and\ \citenamefont
  {Garrahan}(2014)}]{33}%
  \BibitemOpen
  \bibfield  {author} {\bibinfo {author} {\bibfnamefont {I.}~\bibnamefont
  {Lesanovsky}}\ and\ \bibinfo {author} {\bibfnamefont {J.~P.}\ \bibnamefont
  {Garrahan}},\ }\bibfield  {title} {\bibinfo {title} {Out-of-equilibrium
  structures in strongly interacting rydberg gases with dissipation},\ }\href
  {https://doi.org/10.1103/PhysRevA.90.011603} {\bibfield  {journal} {\bibinfo
  {journal} {Phys. Rev. A}\ }\textbf {\bibinfo {volume} {90}},\ \bibinfo
  {pages} {011603} (\bibinfo {year} {2014})}\BibitemShut {NoStop}%
\bibitem [{\citenamefont {Letscher}\ \emph {et~al.}(2017)\citenamefont
  {Letscher}, \citenamefont {Thomas}, \citenamefont {Niederpr\"um},
  \citenamefont {Fleischhauer},\ and\ \citenamefont {Ott}}]{34}%
  \BibitemOpen
  \bibfield  {author} {\bibinfo {author} {\bibfnamefont {F.}~\bibnamefont
  {Letscher}}, \bibinfo {author} {\bibfnamefont {O.}~\bibnamefont {Thomas}},
  \bibinfo {author} {\bibfnamefont {T.}~\bibnamefont {Niederpr\"um}}, \bibinfo
  {author} {\bibfnamefont {M.}~\bibnamefont {Fleischhauer}},\ and\ \bibinfo
  {author} {\bibfnamefont {H.}~\bibnamefont {Ott}},\ }\bibfield  {title}
  {\bibinfo {title} {Bistability versus metastability in driven dissipative
  rydberg gases},\ }\href {https://doi.org/10.1103/PhysRevX.7.021020}
  {\bibfield  {journal} {\bibinfo  {journal} {Phys. Rev. X}\ }\textbf {\bibinfo
  {volume} {7}},\ \bibinfo {pages} {021020} (\bibinfo {year}
  {2017})}\BibitemShut {NoStop}%
\bibitem [{\citenamefont {Yang}\ \emph {et~al.}(2022)\citenamefont {Yang},
  \citenamefont {Xiong}, \citenamefont {Liu},\ and\ \citenamefont
  {Zhang}}]{35}%
  \BibitemOpen
  \bibfield  {author} {\bibinfo {author} {\bibfnamefont {B.}~\bibnamefont
  {Yang}}, \bibinfo {author} {\bibfnamefont {B.}~\bibnamefont {Xiong}},
  \bibinfo {author} {\bibfnamefont {Z.}~\bibnamefont {Liu}},\ and\ \bibinfo
  {author} {\bibfnamefont {B.}~\bibnamefont {Zhang}},\ }\bibfield  {title}
  {\bibinfo {title} {Exploring the effect of decoherence in a two-dimensional
  rydberg system: Application of the time-dependent variational principle},\
  }\href {https://doi.org/10.1103/PhysRevA.105.052225} {\bibfield  {journal}
  {\bibinfo  {journal} {Phys. Rev. A}\ }\textbf {\bibinfo {volume} {105}},\
  \bibinfo {pages} {052225} (\bibinfo {year} {2022})}\BibitemShut {NoStop}%
\bibitem [{\citenamefont {Singer}\ \emph {et~al.}(2004)\citenamefont {Singer},
  \citenamefont {Reetz-Lamour}, \citenamefont {Amthor}, \citenamefont
  {Marcassa},\ and\ \citenamefont {Weidem\"uller}}]{4}%
  \BibitemOpen
  \bibfield  {author} {\bibinfo {author} {\bibfnamefont {K.}~\bibnamefont
  {Singer}}, \bibinfo {author} {\bibfnamefont {M.}~\bibnamefont
  {Reetz-Lamour}}, \bibinfo {author} {\bibfnamefont {T.}~\bibnamefont
  {Amthor}}, \bibinfo {author} {\bibfnamefont {L.~G.}\ \bibnamefont
  {Marcassa}},\ and\ \bibinfo {author} {\bibfnamefont {M.}~\bibnamefont
  {Weidem\"uller}},\ }\bibfield  {title} {\bibinfo {title} {Suppression of
  excitation and spectral broadening induced by interactions in a cold gas of
  rydberg atoms},\ }\href {https://doi.org/10.1103/PhysRevLett.93.163001}
  {\bibfield  {journal} {\bibinfo  {journal} {Phys. Rev. Lett.}\ }\textbf
  {\bibinfo {volume} {93}},\ \bibinfo {pages} {163001} (\bibinfo {year}
  {2004})}\BibitemShut {NoStop}%
\bibitem [{\citenamefont {Tong}\ \emph {et~al.}(2004)\citenamefont {Tong},
  \citenamefont {Farooqi}, \citenamefont {Stanojevic}, \citenamefont
  {Krishnan}, \citenamefont {Zhang}, \citenamefont {C\^ot\'e}, \citenamefont
  {Eyler},\ and\ \citenamefont {Gould}}]{5}%
  \BibitemOpen
  \bibfield  {author} {\bibinfo {author} {\bibfnamefont {D.}~\bibnamefont
  {Tong}}, \bibinfo {author} {\bibfnamefont {S.~M.}\ \bibnamefont {Farooqi}},
  \bibinfo {author} {\bibfnamefont {J.}~\bibnamefont {Stanojevic}}, \bibinfo
  {author} {\bibfnamefont {S.}~\bibnamefont {Krishnan}}, \bibinfo {author}
  {\bibfnamefont {Y.~P.}\ \bibnamefont {Zhang}}, \bibinfo {author}
  {\bibfnamefont {R.}~\bibnamefont {C\^ot\'e}}, \bibinfo {author}
  {\bibfnamefont {E.~E.}\ \bibnamefont {Eyler}},\ and\ \bibinfo {author}
  {\bibfnamefont {P.~L.}\ \bibnamefont {Gould}},\ }\bibfield  {title} {\bibinfo
  {title} {Local blockade of rydberg excitation in an ultracold gas},\ }\href
  {https://doi.org/10.1103/PhysRevLett.93.063001} {\bibfield  {journal}
  {\bibinfo  {journal} {Phys. Rev. Lett.}\ }\textbf {\bibinfo {volume} {93}},\
  \bibinfo {pages} {063001} (\bibinfo {year} {2004})}\BibitemShut {NoStop}%
\bibitem [{\citenamefont {Heidemann}\ \emph {et~al.}(2008)\citenamefont
  {Heidemann}, \citenamefont {Raitzsch}, \citenamefont {Bendkowsky},
  \citenamefont {Butscher}, \citenamefont {L\"ow},\ and\ \citenamefont
  {Pfau}}]{6}%
  \BibitemOpen
  \bibfield  {author} {\bibinfo {author} {\bibfnamefont {R.}~\bibnamefont
  {Heidemann}}, \bibinfo {author} {\bibfnamefont {U.}~\bibnamefont {Raitzsch}},
  \bibinfo {author} {\bibfnamefont {V.}~\bibnamefont {Bendkowsky}}, \bibinfo
  {author} {\bibfnamefont {B.}~\bibnamefont {Butscher}}, \bibinfo {author}
  {\bibfnamefont {R.}~\bibnamefont {L\"ow}},\ and\ \bibinfo {author}
  {\bibfnamefont {T.}~\bibnamefont {Pfau}},\ }\bibfield  {title} {\bibinfo
  {title} {Rydberg excitation of bose-einstein condensates},\ }\href
  {https://doi.org/10.1103/PhysRevLett.100.033601} {\bibfield  {journal}
  {\bibinfo  {journal} {Phys. Rev. Lett.}\ }\textbf {\bibinfo {volume} {100}},\
  \bibinfo {pages} {033601} (\bibinfo {year} {2008})}\BibitemShut {NoStop}%
\bibitem [{\citenamefont {Urban}\ \emph {et~al.}(2009)\citenamefont {Urban},
  \citenamefont {Johnson}, \citenamefont {Henage}, \citenamefont {Isenhower},
  \citenamefont {Yavuz}, \citenamefont {Walker},\ and\ \citenamefont
  {Saffman}}]{7}%
  \BibitemOpen
  \bibfield  {author} {\bibinfo {author} {\bibfnamefont {E.}~\bibnamefont
  {Urban}}, \bibinfo {author} {\bibfnamefont {T.~A.}\ \bibnamefont {Johnson}},
  \bibinfo {author} {\bibfnamefont {T.}~\bibnamefont {Henage}}, \bibinfo
  {author} {\bibfnamefont {L.}~\bibnamefont {Isenhower}}, \bibinfo {author}
  {\bibfnamefont {D.~D.}\ \bibnamefont {Yavuz}}, \bibinfo {author}
  {\bibfnamefont {T.~G.}\ \bibnamefont {Walker}},\ and\ \bibinfo {author}
  {\bibfnamefont {M.}~\bibnamefont {Saffman}},\ }\bibfield  {title} {\bibinfo
  {title} {Observation of rydberg blockade between two atoms},\ }\href
  {https://doi.org/10.1038/NPHYS1178} {\bibfield  {journal} {\bibinfo
  {journal} {Nat. Phys.}\ }\textbf {\bibinfo {volume} {5}},\ \bibinfo {pages}
  {110} (\bibinfo {year} {2009})}\BibitemShut {NoStop}%
\bibitem [{\citenamefont {Gaetan}\ \emph {et~al.}(2009)\citenamefont {Gaetan},
  \citenamefont {Miroshnychenko}, \citenamefont {Wilk}, \citenamefont {Chotia},
  \citenamefont {Viteau}, \citenamefont {Comparat}, \citenamefont {Pillet},
  \citenamefont {Browaeys},\ and\ \citenamefont {Grangier}}]{8}%
  \BibitemOpen
  \bibfield  {author} {\bibinfo {author} {\bibfnamefont {A.}~\bibnamefont
  {Gaetan}}, \bibinfo {author} {\bibfnamefont {Y.}~\bibnamefont
  {Miroshnychenko}}, \bibinfo {author} {\bibfnamefont {T.}~\bibnamefont
  {Wilk}}, \bibinfo {author} {\bibfnamefont {A.}~\bibnamefont {Chotia}},
  \bibinfo {author} {\bibfnamefont {M.}~\bibnamefont {Viteau}}, \bibinfo
  {author} {\bibfnamefont {D.}~\bibnamefont {Comparat}}, \bibinfo {author}
  {\bibfnamefont {P.}~\bibnamefont {Pillet}}, \bibinfo {author} {\bibfnamefont
  {A.}~\bibnamefont {Browaeys}},\ and\ \bibinfo {author} {\bibfnamefont
  {P.}~\bibnamefont {Grangier}},\ }\bibfield  {title} {\bibinfo {title}
  {Observation of collective excitation of two individual atoms in the rydberg
  blockade regime},\ }\href {https://doi.org/10.1038/NPHYS1183} {\bibfield
  {journal} {\bibinfo  {journal} {Nat. Phys.}\ }\textbf {\bibinfo {volume}
  {5}},\ \bibinfo {pages} {115} (\bibinfo {year} {2009})}\BibitemShut {NoStop}%
\bibitem [{\citenamefont {Burrage}\ \emph {et~al.}(2004)\citenamefont
  {Burrage}, \citenamefont {Burrage},\ and\ \citenamefont {Tian}}]{14}%
  \BibitemOpen
  \bibfield  {author} {\bibinfo {author} {\bibfnamefont {K.}~\bibnamefont
  {Burrage}}, \bibinfo {author} {\bibfnamefont {P.}~\bibnamefont {Burrage}},\
  and\ \bibinfo {author} {\bibfnamefont {T.}~\bibnamefont {Tian}},\ }\bibfield
  {title} {\bibinfo {title} {Numerical methods for strong solutions of
  stochastic differential equations: an overview},\ }\href
  {https://doi.org/10.1098/rspa.2003.1247} {\bibfield  {journal} {\bibinfo
  {journal} {Proc. Roy. Soc. Lond. Ser. A}\ }\textbf {\bibinfo {volume}
  {460}},\ \bibinfo {pages} {373} (\bibinfo {year} {2004})}\BibitemShut
  {NoStop}%
\bibitem [{\citenamefont {Zanna}(1999)}]{15}%
  \BibitemOpen
  \bibfield  {author} {\bibinfo {author} {\bibfnamefont {A.}~\bibnamefont
  {Zanna}},\ }\bibfield  {title} {\bibinfo {title} {Collocation and relaxed
  collocation for the fer and the magnus expansions},\ }\href@noop {}
  {\bibfield  {journal} {\bibinfo  {journal} {SIAM J. Numer. Anal.}\ }\textbf
  {\bibinfo {volume} {36}},\ \bibinfo {pages} {1145} (\bibinfo {year}
  {1999})}\BibitemShut {NoStop}%
\bibitem [{\citenamefont {Aparicio}\ \emph {et~al.}(2005)\citenamefont
  {Aparicio}, \citenamefont {Malham},\ and\ \citenamefont {Oliver}}]{16}%
  \BibitemOpen
  \bibfield  {author} {\bibinfo {author} {\bibfnamefont {N.}~\bibnamefont
  {Aparicio}}, \bibinfo {author} {\bibfnamefont {S.}~\bibnamefont {Malham}},\
  and\ \bibinfo {author} {\bibfnamefont {M.}~\bibnamefont {Oliver}},\
  }\bibfield  {title} {\bibinfo {title} {Numerical evaluation of the evans
  function by magnus integration},\ }\href
  {https://doi.org/10.1007/s10543-005-0001-8} {\bibfield  {journal} {\bibinfo
  {journal} {BIT}\ }\textbf {\bibinfo {volume} {45}},\ \bibinfo {pages} {219}
  (\bibinfo {year} {2005})}\BibitemShut {NoStop}%
\bibitem [{\citenamefont {Blanes}\ \emph {et~al.}(2009)\citenamefont {Blanes},
  \citenamefont {Casas}, \citenamefont {Oteo},\ and\ \citenamefont {Ros}}]{1}%
  \BibitemOpen
  \bibfield  {author} {\bibinfo {author} {\bibfnamefont {S.}~\bibnamefont
  {Blanes}}, \bibinfo {author} {\bibfnamefont {F.}~\bibnamefont {Casas}},
  \bibinfo {author} {\bibfnamefont {J.~A.}\ \bibnamefont {Oteo}},\ and\
  \bibinfo {author} {\bibfnamefont {J.}~\bibnamefont {Ros}},\ }\bibfield
  {title} {\bibinfo {title} {The magnus expansion and some of its
  applications},\ }\href {https://doi.org/10.1016/j.physrep.2008.11.001}
  {\bibfield  {journal} {\bibinfo  {journal} {Phys. Rep.}\ }\textbf {\bibinfo
  {volume} {470}},\ \bibinfo {pages} {151} (\bibinfo {year}
  {2009})}\BibitemShut {NoStop}%
\bibitem [{\citenamefont {Tamura}\ \emph {et~al.}(2020)\citenamefont {Tamura},
  \citenamefont {Yamakoshi},\ and\ \citenamefont {Nakagawa}}]{36}%
  \BibitemOpen
  \bibfield  {author} {\bibinfo {author} {\bibfnamefont {H.}~\bibnamefont
  {Tamura}}, \bibinfo {author} {\bibfnamefont {T.}~\bibnamefont {Yamakoshi}},\
  and\ \bibinfo {author} {\bibfnamefont {K.}~\bibnamefont {Nakagawa}},\
  }\bibfield  {title} {\bibinfo {title} {Analysis of coherent dynamics of a
  rydberg-atom quantum simulator},\ }\href
  {https://doi.org/10.1103/PhysRevA.101.043421} {\bibfield  {journal} {\bibinfo
   {journal} {Phys. Rev. A}\ }\textbf {\bibinfo {volume} {101}},\ \bibinfo
  {pages} {043421} (\bibinfo {year} {2020})}\BibitemShut {NoStop}%
\bibitem [{\citenamefont {Samajdar}\ \emph {et~al.}(2021)\citenamefont
  {Samajdar}, \citenamefont {Ho}, \citenamefont {Pichler}, \citenamefont
  {Lukin},\ and\ \citenamefont {Sachdev}}]{37}%
  \BibitemOpen
  \bibfield  {author} {\bibinfo {author} {\bibfnamefont {R.}~\bibnamefont
  {Samajdar}}, \bibinfo {author} {\bibfnamefont {W.~W.}\ \bibnamefont {Ho}},
  \bibinfo {author} {\bibfnamefont {H.}~\bibnamefont {Pichler}}, \bibinfo
  {author} {\bibfnamefont {M.~D.}\ \bibnamefont {Lukin}},\ and\ \bibinfo
  {author} {\bibfnamefont {S.}~\bibnamefont {Sachdev}},\ }\bibfield  {title}
  {\bibinfo {title} {Quantum phases of rydberg atoms on a kagome lattice},\
  }\href {https://doi.org/10.1073/pnas.2015785118} {\bibfield  {journal}
  {\bibinfo  {journal} {Proc. Natl. Acad. Sci.}\ }\textbf {\bibinfo {volume}
  {118}},\ \bibinfo {pages} {e2015785118} (\bibinfo {year} {2021})}\BibitemShut
  {NoStop}%
\bibitem [{\citenamefont {Semeghini}\ \emph {et~al.}(2021)\citenamefont
  {Semeghini}, \citenamefont {Levine}, \citenamefont {Keesling}, \citenamefont
  {Ebadi}, \citenamefont {Wang}, \citenamefont {Bluvstein}, \citenamefont
  {Verresen}, \citenamefont {Pichler}, \citenamefont {Kalinowski},
  \citenamefont {Samajdar} \emph {et~al.}}]{38}%
  \BibitemOpen
  \bibfield  {author} {\bibinfo {author} {\bibfnamefont {G.}~\bibnamefont
  {Semeghini}}, \bibinfo {author} {\bibfnamefont {H.}~\bibnamefont {Levine}},
  \bibinfo {author} {\bibfnamefont {A.}~\bibnamefont {Keesling}}, \bibinfo
  {author} {\bibfnamefont {S.}~\bibnamefont {Ebadi}}, \bibinfo {author}
  {\bibfnamefont {T.~T.}\ \bibnamefont {Wang}}, \bibinfo {author}
  {\bibfnamefont {D.}~\bibnamefont {Bluvstein}}, \bibinfo {author}
  {\bibfnamefont {R.}~\bibnamefont {Verresen}}, \bibinfo {author}
  {\bibfnamefont {H.}~\bibnamefont {Pichler}}, \bibinfo {author} {\bibfnamefont
  {M.}~\bibnamefont {Kalinowski}}, \bibinfo {author} {\bibfnamefont
  {R.}~\bibnamefont {Samajdar}}, \emph {et~al.},\ }\bibfield  {title} {\bibinfo
  {title} {Probing topological spin liquids on a programmable quantum
  simulator},\ }\href {https://doi.org/10.1126/science.abi8794} {\bibfield
  {journal} {\bibinfo  {journal} {Science}\ }\textbf {\bibinfo {volume}
  {374}},\ \bibinfo {pages} {1242} (\bibinfo {year} {2021})}\BibitemShut
  {NoStop}%
\end{thebibliography}%

\end{document}